# Characterizations of thermal stability and electrical performance of Au-Ni coating on CuCrZr substrate for high vacuum radio-frequency contact application

Z.X. Chen[1,2*], J. Hillairet[1], V. Turq[2], Y.T. Song[3], R. Laloo[2], J.M. Bernard[1], K. Vulliez[4], G. Lombard[1], C. Hernandez[1], Q.X. Yang[3], L. Ferreira[5], F. Fesquet[5], P. Mollard[1], R. Volpe[1]

[1] CEA, IRFM, F-13108 Saint-Paul-Lez-Durance, France
[2] Institut Carnot CIRIMAT, UMR CNRS-UPS-INP 5085, Université Paul-Sabatier, 118 route de Narbonne, 31062 Toulouse cedex 9, France
[3] Institute of Plasma Physics, CAS, Hefei, Anhui 230031, China
[4] Laboratoire d'étanchéité, DEN/DTEC/SDTC, CEA, 2 rue James Watt 26700 Pierrelatte, France
[5] CERN, Geneva, Switzerland

**Abstract**

Radio-frequency (RF) contacts—which are an example of electrical contacts—are commonly employed on accelerators and nuclear fusion experimental devices. RF contacts with a current load of 2 kA for steady-state operation were designed for application to the International Thermonuclear Experimental Reactor (ITER) device. In contrast to the typical working conditions of general commercial electrical contacts, those of RF contacts employed on fusion devices include high vacuum, high temperature, and neutron radiation. CuCrZr is currently of interest as a base material for the manufacture of louvers of RF contacts, which has excellent thermal and electrical properties and has low creep rate at 250 °C. In this study, a hard Au coating (Au-Ni) was electroplated on CuCrZr samples and the samples were then subjected to thermal aging treatment at 250 °C for 500 h in order to simulate the vacuum-commissioning process of the ITER. The effects of thermal aging on the hardness, elastic modulus, crystallite size, and compositions of the coating were investigated via microstructural and mechanical characterizations of the coating material. Metal atom migration in different coating layers during thermal aging was characterized and evaluated via scanning electron microscopy/energy dispersive X-ray spectroscopy observations of the cross-sectional surfaces, and the obtained results could be used to directly select the coating thickness for the final RF contact component. The contact resistance—an important parameter of the RF contact—was measured in a dedicated testbed built to simulate fusion reactor conditions between CuCrZr pins and stainless steel plates coated with Au-Ni and Rh, respectively.

**Keyword:** Au-Ni coating, thermal aging, diffusion, contact resistance

## 1. Introduction

Electrical contacts that operate under alternating current (AC), e.g., radio-frequency (RF) contacts, are commonly employed on large-scale scientific research devices in several fields such as high-energy physics and nuclear fusion [1, 2]. Electrical contacts employed on accelerators or nuclear-fusion-related research devices differ from general commercial electrical contacts mainly in terms of the specific working conditions: electrical contacts operate under conditions of long periods of high-temperature (250 °C) baking to attain ultrahigh vacuum and are subjected to high heat loads owing to ohmic losses during operation.

Moreover, RF contacts—as flexible electrical connections between RF conductors—must have the ability to absorb linear movement produced by thermal expansion of the RF conductors. Sliding in high vacuum makes selection of the material of the RF contact more challenging since severe wear can occur on the RF contact. An RF contact that can be operated safely for 1200 s with a maximum current load of 2 kA under high-vacuum and high-temperature conditions is presently under development at the Commissariat à l'Énergie Atomique (CEA) in France for the International Thermonuclear Experimental Reactor (ITER) project, which is the world's largest fusion experimental facility [1]. The first prototype of an RF contact whose louvers were made of Cu with a 4 μm Au coating was tested. The test revealed that the thin Au coating on the louvers was worn out easily. Moreover, under high-temperature conditions, the strength of pure Cu was insufficient and the creep was significant.

The CuCrZr alloy is widely used as a heat sink material in fusion experimental devices [3-7] because of its excellent thermal conductivity, strength, and fatigue resistance [8]. In addition to having good mechanical performance, CuCrZr can also exhibit an electrical conductivity of about 77% IACS (International Annealed Copper Standard) at room temperature [9, 10]. Therefore, CuCrZr has attracted interest for application as a material for the manufacture of ITER RF contacts. As is the case with pure Cu, when CuCrZr is exposed to an $O_2$-containing atmosphere, $Cu_2O$, CuO, or both will form on the CuCrZr surface because of oxidation [8, 11-13]. An Au layer, as a functional coating for oxidation protection as well as for minimization of the contact resistance, is applied on the surface of RF contact ouvers made of CuCrZr. The RF contact will slide against the RF conductors. One of the candidate materials for the manufacture of RF conductors is 316L stainless steel coated with Rh. However, the wear resistance of pure Au, which is a soft material, is poor [14], and it could worsen under conditions of high vacuum and high temperature.

Hard Au containing a small amount Ni or Co can have significantly higher hardness than pure Au [15-19] and consequently exhibit higher wear performance. Replacement of a pure Au coating with a hard Au coating may increase the working reliability and prolong the lifetime of an RF contact. In view of nuclear safety considerations, Co is a strictly controlled impurity in ITER materials, which can be activated easily by neutron radiation from a fusion plasma [20, 21]; therefore, application of a Au-Ni coating is preferable. However, few studies have been conducted on the application of the Au-Ni coating on a CuCrZr alloy, and no studies have evaluated the Au-Ni coating working under ITER relevant working conditions.

## 2. Materials and experiments

### 2.1 Materials and specimen preparation

Plate samples with dimensions of 10 mm × 8 mm × 2 mm and pin samples with a length of 25 mm, diameter of 5 mm, and a curved tip surface with a radius of 8 mm were prepared using CuCrZr (ASTM C18150: 0.7 wt.% Cr, 0.04 wt.% Zr, and balance Cu). The values of their surface roughness $Sa$ (arithmetic mean height) were 0.44 μm and 1.6 μm, respectively.

The main functional coating for improving the performance of an electrical contact was Au-Ni, which was expected to be electroplated on the CuCrZr samples. A previous study [22] reported that after baking at 250 °C for 1 month, 2.5 μm thick of fine gold interdiffused with copper from the substrate. Since ITER RF contacts are expected to be used under high-temperature, high-vacuum conditions with intensive sliding, application of a pure hard Ni interlayer as a diffusion barrier between Cu and Au atoms is necessary. In order to reduce the effects of diffusion and wear on the electrical performance of the RF contact, a 15-µm-thick Au-Ni coating was planned to be electroplated on the actual RF contact component. In the preliminary research stage of materials for RF contacts, a 3.4-µm-thick Au-Ni coating with a 4.5-µm-thick Ni interlayer was electrodeposited on the plate samples, which were used for characterization of the surface roughness, porosity, cracking, hardness, grain size, and metal diffusion. Further, a 15-µm-thick Au-Ni coating was electroplated on the pin samples—which were used for contact resistance measurements—in order to obtain realistic results that could be applied directly to the ITER.

The diffusion phenomenon, which occurs during long-term high-temperature thermal aging in vacuum, and its effects should not be neglected, since they may alter the mechanical and electrical properties of the coating. In order to simulate the ITER baking process, thermal aging treatment was performed on a group of coated samples under ITER-specific baking conditions (i.e., high-vacuum and high-temperature conditions of $10^{-6}$ Pa and 250 °C, respectively). These samples were heated from room temperature to 250 °C at a heating rate of 1.3 °C/min and then aged at 250 °C for 500 h. The samples were subsequently were cooled down to room temperature at a rate of 0.16 °C/min under vacuum.

2.2 Microstructural and mechanical characterizations

The mechanical properties of the coating were characterized at room temperature by using a CSM® ultra nanoindentation tester (CSEM, Switzerland) with force and displacement resolutions of 1 nN and 0.0003 nm, respectively. The morphology of the samples was observed by scanning electron microscopy (SEM) coupled with energy-dispersive X-ray spectroscopy (EDS) using the JEOL JSM-6510LV electron microscope (acceleration voltage of 20 kV). Two views were selected for analysis: the top view was selected to observe the coating quality and crack spacing, and the cross-sectional view was selected to measure the coating thickness and crack depth. A morphology study of the coating surface was conducted to investigate the effects of high-temperature thermal aging and solid diffusion on the Au-Ni coating. Three-dimensional (3D) surface measurements were performed on a 3D optical profiler (Sensofar, USA) to determine the surface roughness of the coating. Crystal structure information of the Au-Ni coating, including the crystal phase, crystallite size, and lattice constants, was acquired by X-ray diffraction (XRD) using the D4 Endeavor diffractometer (Bruker, Germany) with a Ni filter and Cu Kα radiation ($\lambda$ = 1.54184 Å, 40 kV, 40 mA). The XRD data were collected in the 2θ range of 10°–100° with a step scan of 0.0157°.

2.3 Measurement of electrical contact resistance

The wear performance of Au coating can be improved by alloying with Ni. However, this alloying causes a reduction in the electrical conductivity of the Au coating. Microstructure and hardness are important factors influencing the contact performance of the Au-Ni coating. The microstructural changes of a material after thermal aging cause corresponding changes in its electrical and mechanical properties. Such changes in properties would, in turn, alter the electrical performance of the initially designed electrical contact. For the ITER RF contact, under a current load of 2 kA, an increase of 1 mΩ in the contact resistance can impose a heat load of 4 kW on the component. This heat load is a large burden on the mechanical structure of the RF contact, which is designed for steady-state operation. Therefore, the contact resistance should be investigated in detail on a dedicated testbed.

## 3. Results and discussion

3.1 SEM observations and interferometric study of surface morphology

The microstructure of the surface of the original Au-Ni coating was observed under ×50000 magnification, as shown in Fig. 1. A grain structure with cauliflower-like nodules was observed, with an average grain size of around 20 nm. No coating defects such as cracks and pores were observed in the SEM image of the coating. After thermal aging, no obvious increase in the grain size was observed in the SEM image. However, numerous pores (number density: 1.7/µm$^2$) with an average diameter of 48 nm were formed on the surface of the coating. During the long period of thermal aging, recovery, recrystallization, and grain growth occurred in the Au-Ni coating. Boundary migration occurred simultaneously, and these pores were probably formed among the grain boundaries owing to discontinuities of boundary movement [23, 24]. As these pores were extremely small in size, they did not have any obvious effects on the electrical and tribological performances of the coating.

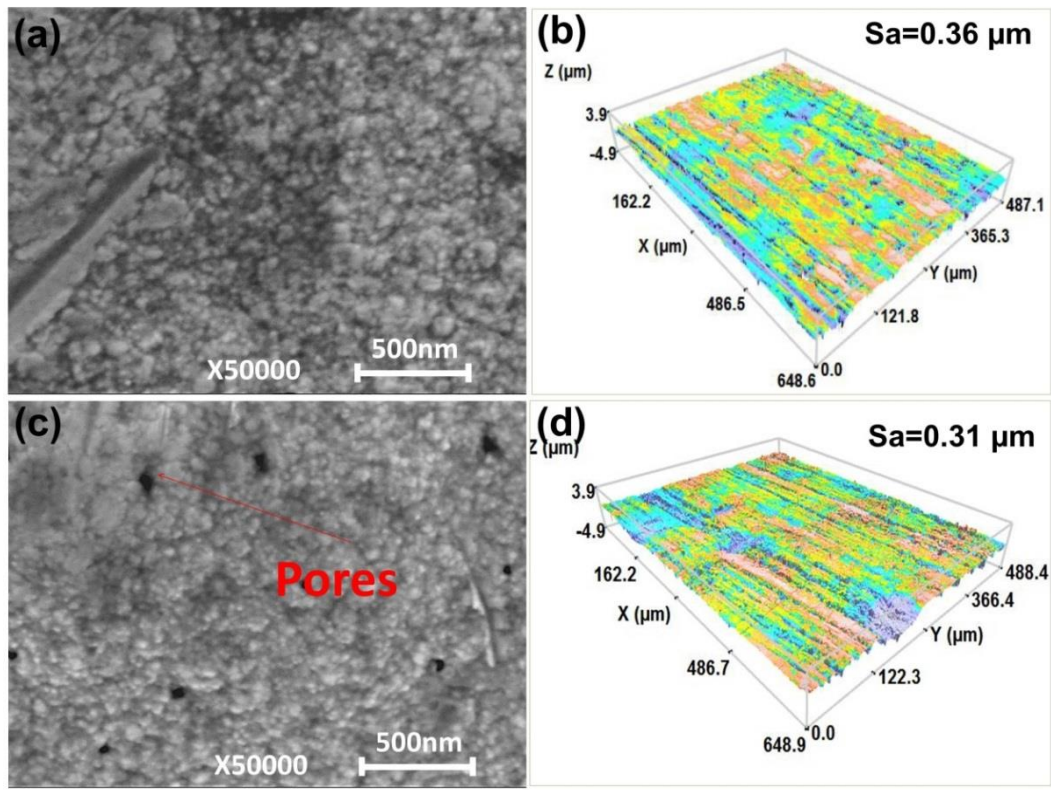

**Fig. 1 Observation results of surface morphology of coating surfaces: (a) SEM image of surface of original coating, (b) 3D interferometric image of surface of original coating, (c) SEM image of surface of thermally aged coating, (d) 3D interferometric image of surface of thermally aged coating**

The surface topography of the Au-Ni coating on the CuCrZr substrate was also studied. The *Sa* value of the original coating was about 0.36 µm, which indicated that the original Au-Ni coating was smooth. After thermal aging, there was no obvious change in the surface roughness of the coating, and *Sa* of the thermally aged coating was 0.31 μm.

3.2 XRD study of grain size and lattice parameter

The crystallite size of a coating is closely related to the coating's mechanical properties, e.g., its hardness, which would affect the wear performance of the coating. The crystallite size of a coating can be calculated using the Scherrer equation, which is expressed as [25, 26]:

$$D = \frac{K \cdot \lambda}{\beta \cdot \cos\theta}$$

where $D$ is the mean size of the crystallites, which is equal to or smaller than the grain size of the material; $\theta$ is the Bragg angle; $\beta$ is the full-width at half-maximum (FWHM) of a diffraction line located at $\theta$ which is subtracted from the peak broadening caused by the facility itself [27]; $\lambda$ is the X-ray wavelength (Cu Kα radiation with $\lambda = 0.15418$ nm); and $K$ is the Scherrer constant, which depends on the peak breadth, crystallite shape, and crystallite size distribution. In this study, $K$ was selected as 0.9 [28-31].

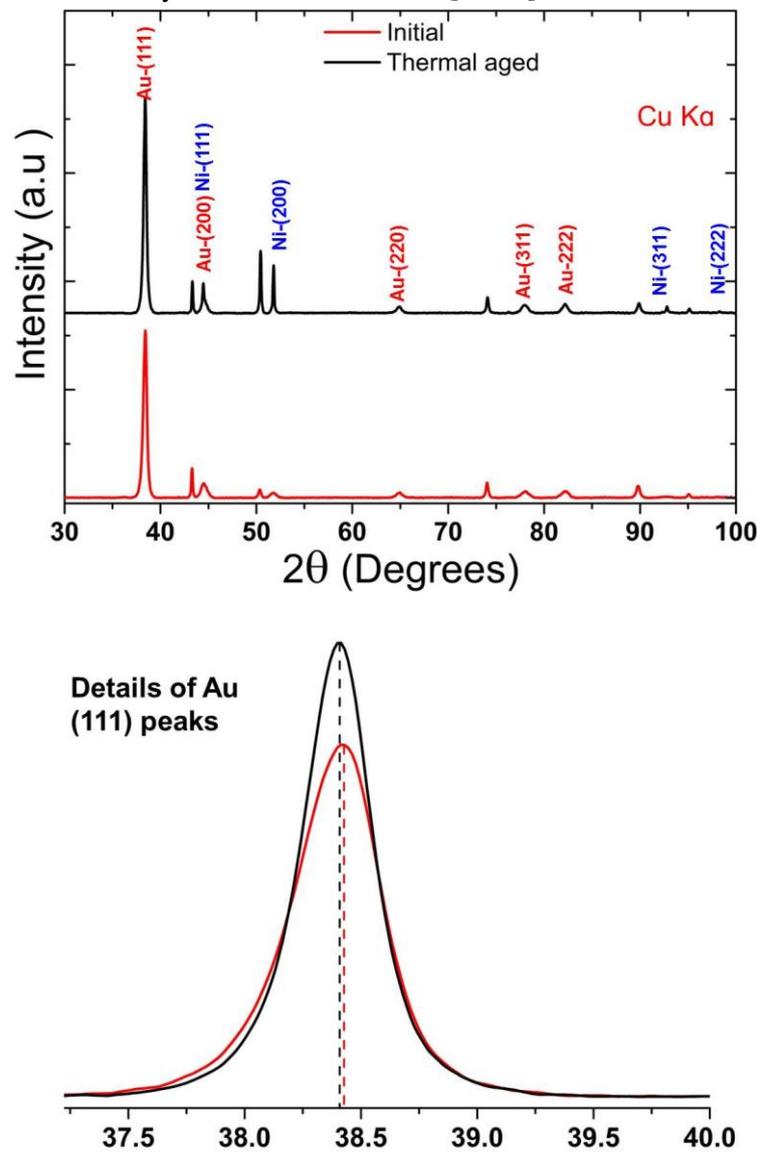

**Fig. 2 XRD pattern of Au-Ni coating on CuCrZr (original and thermally aged)**

As shown in Fig. 2, before thermal aging, peaks of the Au, Cu, and Ni phases are present in the XRD pattern. The Ni peaks originate from the Ni interlayer of the Au-Ni coating and the Cu peaks originate from the substrate. From the (111) peak and according to the Scherrer formula, the crystallite size is determined to be about 20.1 nm, with a lattice constant of 0.4058 nm. The right shift of the Au peaks and the smaller lattice constant than that of pure Au were caused by the co-deposition of the Ni element, having a smaller atomic diameter, into the Au layer; the higher the substitution of Au by Ni, the smaller the Au lattice becomes. The Au (111) peak shifted by 0.03° in the low-$\theta$ direction, which may have been caused by the diffusion of Cu atoms (whose atomic radius is larger than that of Ni) into the Au-Ni layer. In comparison to the XRD pattern of the initial sample, after thermal aging, the FWHM of the Au peaks did not decrease noticeably, which means that the Au-Ni coating remained fairly unchanged during thermal aging, without severe grain coarsening. The crystallite size was calculated from the Au (111) peak to be about 24.4 nm. The lattice constant of the Au–Ni coating after thermal aging was 0.4062 nm. Grain coarsening did not have any significant effect on the hardness of the coating. The crystallite size of the Ni interlayer was calculated from the Ni (200) peak. Obvious grain coarsening of the Ni interlayer was observed, which indicated that after the long-term thermal aging, the crystallite size of the Ni interlayer increased from 14.4 nm to more than 100 nm.

3.2 Study of mechanical properties

A three-sided pyramidal Berkovich diamond indenter was used to indent the Au-Ni coatings under different treatments; the Young's modulus and Poisson's ratio of the indenter tip were 1140 GPa and 0.07, respectively [32]. The CuCrZr substrate and Ni interlayer may influence the measured mechanical properties, especially the elastic modulus, during the indentation of the Au-Ni coatings; therefore, the penetration depth should be kept smaller than 1/10th of the thickness of the Au-Ni coatings [33]. In view of this requirement of the penetration depth, a peak load of 7 mN was applied. In all cases, under the peak load of 7 mN, the loading rate and unloading rate were set to the same value of 14 mN/min. During the nanoindentation tests, when the loads reached 7 mN, the maximum loads were maintained for 60 s to observe the creep of the coatings. The obtained load–displacement curves were analyzed by the standard Oliver–Pharr method to determine the elastic modulus $E$ and hardness $H$ of the coatings [34-36]. Then, the Vickers hardness of the coatings was calculated using the obtained indentation hardness (HV ≈ $H$/10.8). The Poisson's ratio of the Au-Ni coatings was assumed to be 0.3. In order to enhance the reproducibility of the measured results, in each sample, measurements were performed 10 times at each of the different selected positions.

From the load–displacement curves in Fig. 3, it can be seen that the results of the 10 measurements on the original coating, especially during loading, were more homogenous than those on the thermally aged coating. It has been reported that differences in surface roughness lead to dispersion of loading curves [37]. However, in this study, the original and thermally aged coatings had similar surface roughness values, and so, the large dispersion of the load–displacement curves for the thermally aged coating was caused by other factors. Physical (thickness) and chemical (composition) changes that occurred in the Au-Ni coating after thermal aging may be the possible explanations of this dispersion. In addition to the large scattering of the residual indentation depths for the thermally aged coating as obtained from its load–displacement curves, another obvious difference from the original coating was observed: a decreased residual indentation depth. A smaller imprint depth indicates that the hardness of the thermally aged coating increased. In addition, during the load hold period, the thermally aged coating showed better creep resistance than the original coating because of the smaller creep deformation of the former. The elastic modulus and hardness of the original Au-Ni coating were calculated to be 133.3 ± 4.6 GPa and 267.7 ± 19.0 HV, respectively. These values changed to 135 ± 8.0 GPa and 371.2 ± 36.1 HV, respectively, after thermal aging.

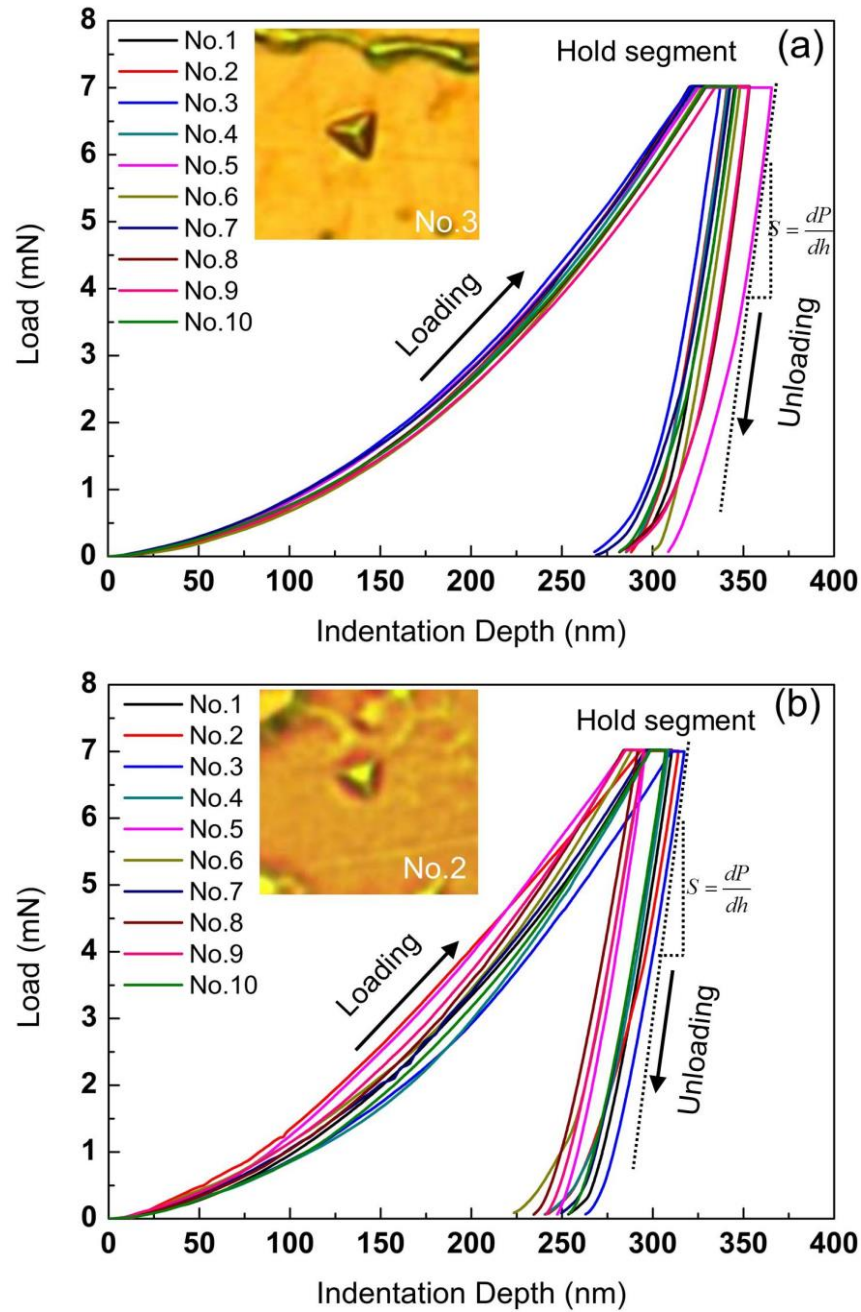

**Fig. 3 Nanoindentation load–displacement curves of Au-Ni coatings: (a) original coating, (b) thermally aged coating**

### 3.4 SEM-EDS study of diffusion

Changes in the material properties and coating heterogeneity on the thermally aged Au-Ni coating were observed by the nanoindentation tests. The reasons for the occurrence of these phenomena were investigated by SEM-EDS. Specifically, EDS area mapping and EDS line were performed at the coating interface on the cross-sections of the polished samples. Further, the phenomenon of material interpenetration between different coating layers due to diffusion was observed. Quantitative EDS point analyses were performed to determine the coating compositions.

As shown in Fig. 4, in the case of the original coating, the interfaces between Au-Ni/Ni, and CuCrZr/Ni substrate were clearly observed with high contrast, and no obvious broad transition layers between the coating layers and their alloy substrate were observed. The element depth profiles showed the presence of a small amount of Cu, in addition to Ni, in the Au-Ni coating. The original thickness of the Au-Ni coating was about 3.5 μm, and a 4.3-μm-thick Ni interlayer was applied, as mentioned earlier. However, after thermal aging, considerable diffusion occurred at the CuCrZr/Ni interface and the Ni/Au-Ni interface. Obvious penetration of Ni into the Au-Ni layer was observed at different locations, which caused blurring of the bonding interface and a decrease in the thickness of the Au-Ni coating. Through statistical analysis, the remaining thicknesses of the Au-Ni and Ni layers were determined to be $1.84 \pm 0.46$ μm and $3.04 \pm 0.42$ μm, respectively. The considerable diffusion of Ni into the Au-Ni layer can be explained as follows: during the thermal aging at 250 °C for 500 h, the migration of the Ni atoms from the Ni interlayer to the Au-Ni layer was accelerated and the oversaturated Ni soluted in the Au matrix and precipitated from this matrix to form a Ni-rich phase. In some areas, the Au-Ni layer had already disappeared and subsequently appeared on the top surface. Interdiffusion between the Ni interlayer and the CuCrZr substrate was also observed, and many voids (Kirkendall voids) were formed because of this interdiffusion (Fig. 4(g) and Fig. 4(h)). The occurrence of this phenomenon was also reflected in the element depth profiles, which showed broadening of the transition layers of Ni and Cu after thermal aging.

The obvious diffusion of Ni into the Au-Ni layer could be one of the reasons for the increase in the hardness of the Au-Ni coating after thermal aging via two hardening mechanisms: solid-solution hardening and precipitation hardening. Moreover, the thickness of the Au-Ni layer decreased in several local areas after the thermal aging process; therefore, the increase in the hardness of the Au-Ni layer after thermal aging was probably due to the effects caused by the Ni interlayer, which has similar hardness values, as reported previously [38].

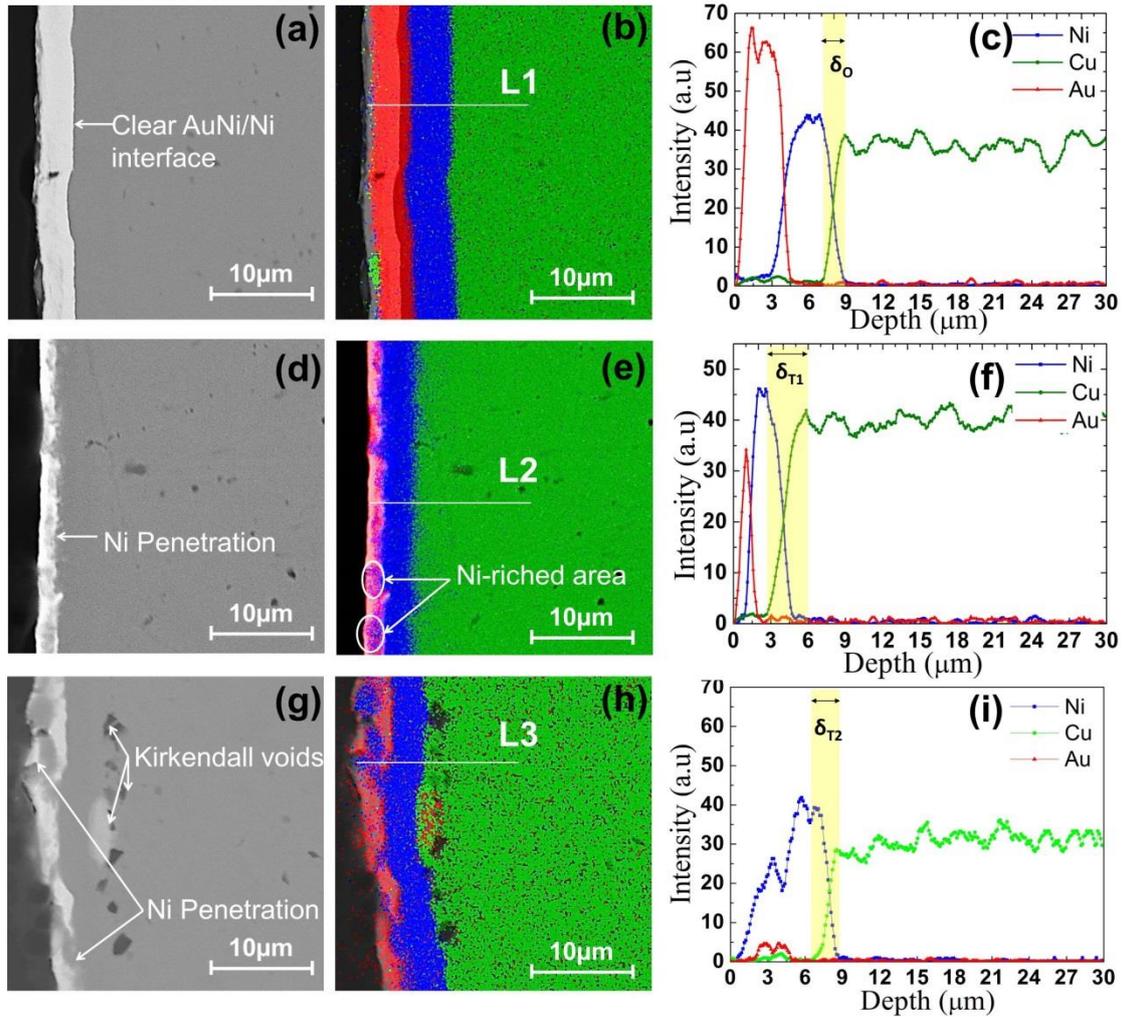

**Fig. 4 Results of SEM-EDS mapping and line scan at coating interfaces: (a, d, g) SEM images of original coating, thermally aged coating at position 1, and thermally aged coating at position 2; (b, e, h) EDS mapping images of original coating, thermally aged coating at position 1, and thermally aged coating at position 2; (c, f, i) element profiles along lines 1, 2, and 3**

Quantitative EDS point analyses were performed, and the diffusion phenomenon was studied via analysis of the chemical compositions of the coating surfaces. As shown in Fig. 4, before thermal aging, the composition was 2.09 wt.% Ni, 1.59 wt.% Cu, and 96.32 wt.% Au. The 2.09 wt.% Ni that alloyed into the coating increased the hardness of the Au matrix significantly. After thermal aging in high vacuum, the Ni content on the coating surface increased to 3.24 wt.% owing to metal diffusion. At a local position (position C in Fig. 5(e)), a sharp increase in the Cu content from 1.59 wt.% to 46.35 wt.% was observed, which indicated that the CuCrZr substrate had invaded into the Au-Ni coating and became exposed to the surroundings without the protection of the coating. Exposure of the Cu present on the Au-Ni coating surface can degrade the corrosion resistance of the RF contacts when they are exposed to air during assembly and maintenance. Direct contact between Cu and Au can result in the formation of intermetallic compounds. Consequently, the electrical and tribological performances of the Au-Ni coating were degraded.

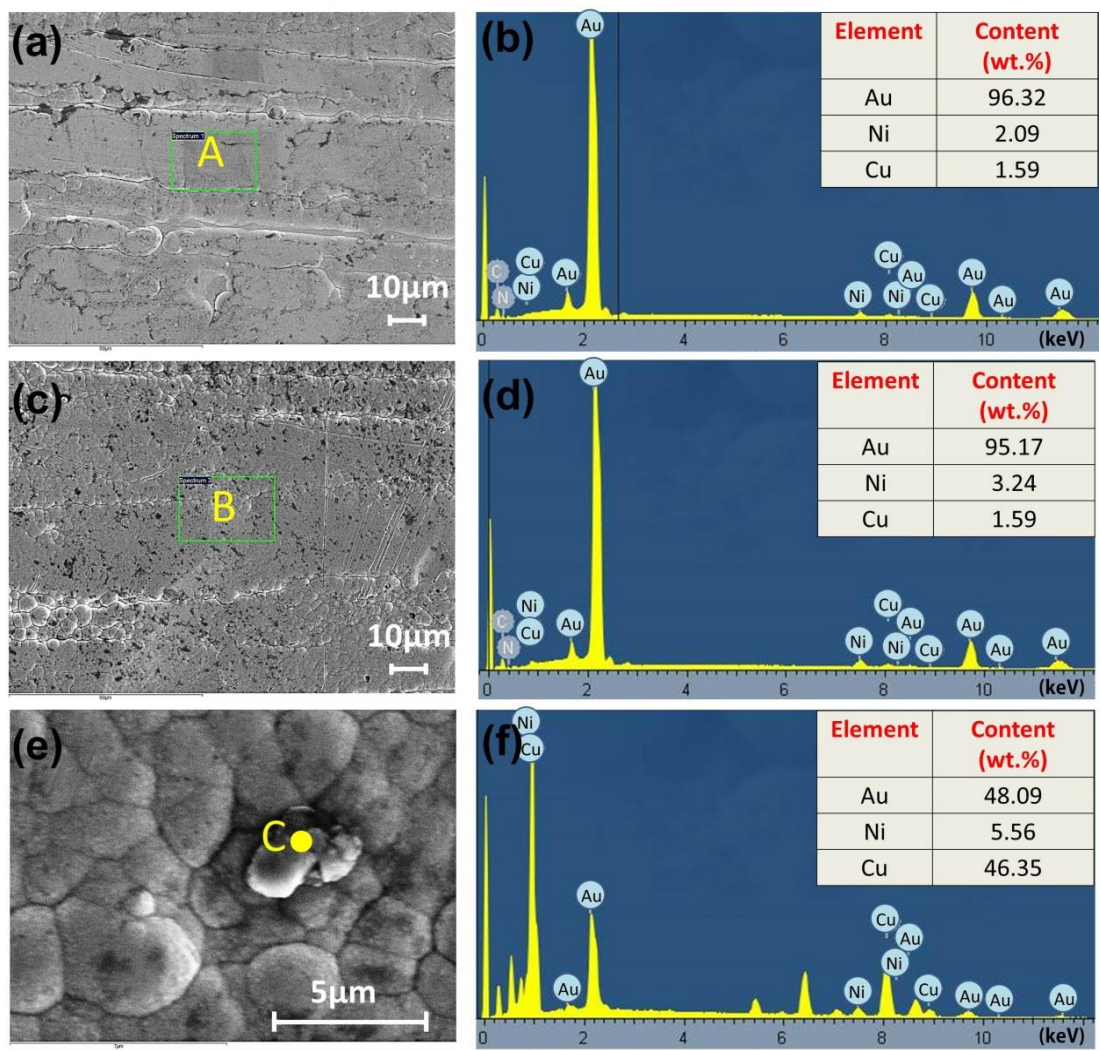

**Fig. 5** SEM/EDS analyses of Au-Ni coating surfaces: (a, c, e) SEM images of surfaces of original Au-Ni coating, thermally aged Au-Ni coating (×1000), and thermally aged Au-Ni coating (×8000), respectively; (b, d, f) EDS spectra of area A, area B, and point C in (a), (c), and (e), respectively

## 4. Effects of operation temperature, vacuum, and normal contact force on contact resistance

When two solid bodies are in contact, the actual contact area that is created under the application of mechanical loads is only a small fraction of the nominal contact area. This actual contact area is affected by mechanical loads, properties of the materials in contact, and their degradation characteristics. Chemical and physical absorption can cause contamination of the contact surfaces and the subsequent formation of a thin insulation layer. Thus, electrical contact between two engineering bodies only occurs at the spot areas randomly distributed on the surfaces when the surface asperities break the insulation films and penetrate into the other contact body (Fig. 6). Contact resistance is an important property of electrical contacts, especially for electrical contacts subject to heavy loads, which induce power loss and cause heating of the electrical contacts. High temperature can cause degradation of material properties, such as an increase in resistivity and acceleration of the creep of the material. Severe creep of the electrical contact material might cause a decrease in the normal contact force and eventually lead to failure. The contact resistance of an ITER RF contact should be controlled to be lower than 7 mΩ.

The electrical performance of the Au-Ni coating was carefully investigated by measurement of the contact resistance (under direct current) between the Au-Ni-coated pin samples and the Rh-coated 316L plates (thickness: ~4.2 μm, surface roughness $Sa$: 1.9 μm) on a dedicated testbed [39]. The pin and plate samples can accurately represent the louver of the RF contact and the RF conductor to which the RF contact is connected, respectively. As shown in Fig. 6, a four-terminal resistance measurement method was used to measure the contact resistance between sample pairs (i.e., pin versus plate). The sample pair was first mounted under atmosphere at room temperature and the contact resistance under various normal contact forces was measured. Then, the environment was pumped to high vacuum ($10^{-3}$–$10^{-4}$ Pa), and the contact resistance under the various normal contact forces was measured again. Finally, the sample pair was heated between 100 °C and 200 °C in steps of 50 °C, and the contact resistance under the various normal contact forces was measured at each temperature.

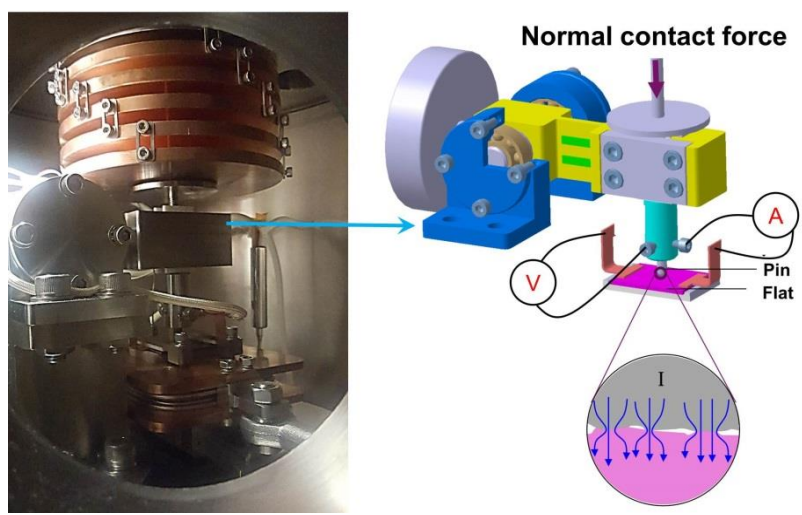

**Fig. 6 Configuration of contact resistance testbed design based on four-terminal resistance measurement method**

Fig. 7 shows the results of the contact resistance measurements of the original and thermally aged sample pairs, respectively. For the original sample pair, the contact resistances in the vacuum environment were smaller than those measured under atmosphere; this trend was more significant at low normal contact forces. The reason for this result is that under atmosphere, the surfaces of the pin and plate samples were covered with a contamination layer of gas and grease owing to physical absorption. This layer acted as electrical insulation during the contact resistance measurements, and it could be removed by vacuum condition and by normal contact force applying. A similar trend of the contact resistances was observed for the thermally aged sample pair. However, the difference between the contact resistances in the vacuum environment and under atmosphere was smaller for the thermally aged sample pair than for the original sample pair, since the contact surfaces of thermally aged sample pair were cleaner. For both the original sample pair and the thermally aged sample pair, an increase in the normal contact force resulted in a decrease in contact resistance. When the normal contact forces were higher than 18 N, the application of a high normal contact force was less effective in achieving a lower contact resistance. The variation trend of the contact resistance with the increase in the test temperature for the thermally aged sample pair was very different from that for the original sample pair. For the original sample pair, the contact resistance decreased continuously with an increase in the test temperature from 25 °C to 200 °C. However, for the thermally aged sample pair, the contact resistance initially increased with an increase in the test temperature and the contact resistance attained a threshold value between the operation temperatures of 150 °C and 200 °C, beyond which the relation between the contact resistance and the test temperature changed.

The effect of increasing temperature on the contact resistance can be explained by two different mechanisms: the increase in the resistivity of the coating materials and the degradation of their mechanical properties. The original Au-Ni coating has a lower elastic modulus and lower hardness; therefore, when the normal contact force was applied to it, it underwent deformation easily. Compared with the resistivity increase, the contact area increase due to the deformation became the predominant factor that affected the contact resistance. As revealed by the nanoindentation tests, the elastic modulus and hardness of the Au-Ni coating increased after thermal aging, and as a result, the coating showed higher thermal stability to withstand deformation. Therefore, unlike in the case of the original Au-Ni coating, the resistivity increase was the main cause for the increase in the contact resistance of the thermally aged coating at operation temperatures lower than 150 °C. Further, significant deformation of the coating occurred at 200 °C, which decreased the contact resistance. These results showed that under almost all the applied normal contact forces, the measurement uncertainties for the thermally aged sample pair were higher than those for the original sample pair, which can be explained by the characterization results of the former, wherein after thermal aging, the material heterogeneity increased because of material diffusion.

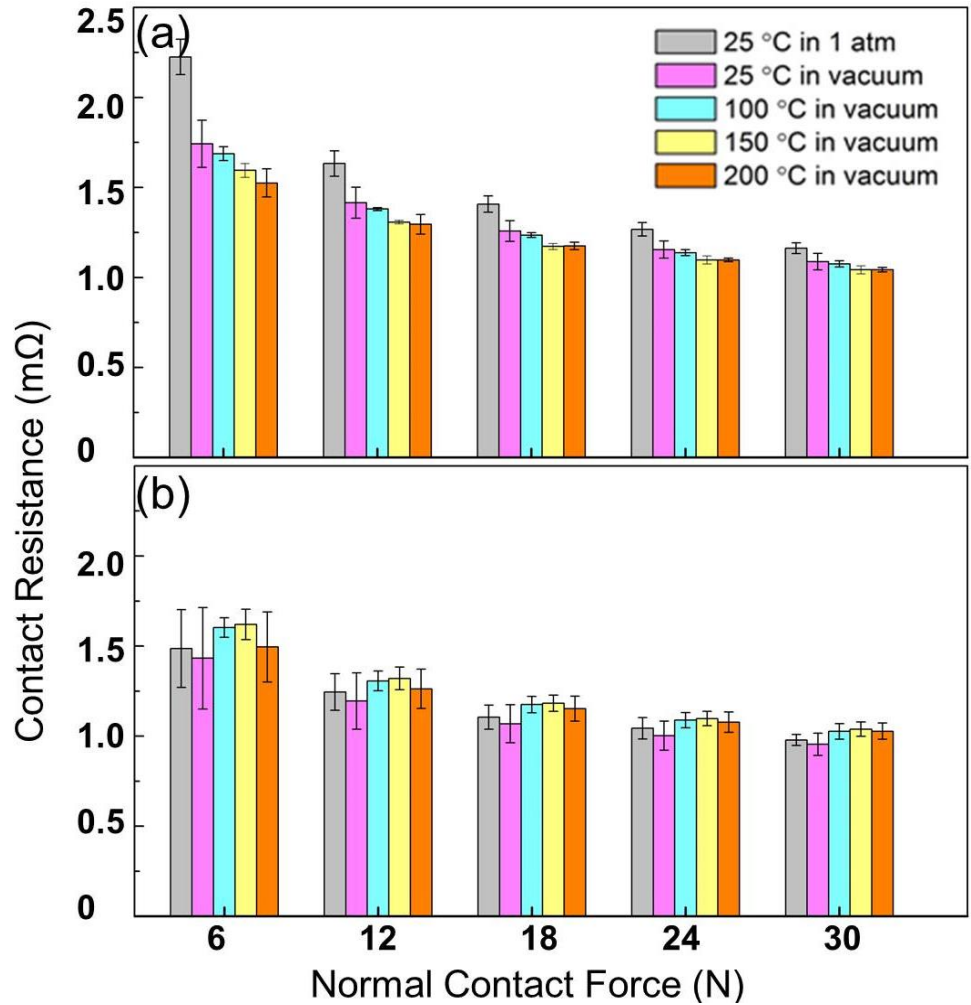

**Fig. 7 Results of contact resistance measurements of Au-Ni/Rh coating pairs: (a) original, (b) thermally aged**

The samples were subjected to thermal aging treatment in order to mimic the condition of high-temperature baking for long periods for attainment of ultrahigh vacuum. Therefore, the thermally aged samples can well reflect the realistic operational behavior of RF contacts. Although a decreasing trend of the contact resistance with increasing normal contact force was observed, it would be preferable to keep the normal contact force below 18 N to achieve a tradeoff between the electrical and tribological performances of the RF contact. Below a normal contact force of 18 N, the contact resistance was around 1.2 mΩ, and this value did not change noticeably with an increase in the operation temperature from 25 °C to 200 °C. In other words, below a normal contact force of 18 N, the effects of operation temperature on the contact resistance of the RF contact can be ignored.

## 5. Conclusion

The development of RF contacts that can meet the operation specifications of the ITER (current load of 2 kA during steady-state operation for 3600 s) is a highly challenging task. To ensure good mechanical strength of RF contact louvers, CuCrZr was selected as the base material for their manufacture. A Au-Ni coating was applied on the surfaces of the RF contact louvers in order to minimize the contact resistance. The feasibility of this material pair for application to ITER RF contacts was evaluated via characterization of the coating after its thermal aging at 250 °C for 500 h in vacuum.

Pores with an average diameter of 48 nm were observed on the surface of the thermally aged Au-Ni coating. However, these pores were not expected to have any significant effects on the effectiveness of the coating. The crystallite size of the thermally aged Au-Ni coating increased from 20.1 nm to 24.4 nm, which showed that the Au-Ni coating has good performance in grain size stability. At the same normal contact force, the contact resistances of the thermally aged sample pair were smaller than those of the original sample pair. However, at a normal contact force higher than 18 N, this difference in contact resistances became smaller and the contact resistances stabilized at relatively low values (around 1.2 mΩ). All these results indicate that Au-Ni is a suitable coating material for ITER RF contacts because it meets the design requirement of their contact resistances.

However, severe invasion of Ni into the 3.5-μm-thick Au-Ni layer was caused by metal interdiffusion, and the Cu phase was observed in some areas on the top surface of the Au-Ni coating. This diffusion phenomenon may degrade the performance of the Au-Ni coating, e.g., its oxidation resistance and wear resistance. Solutions such as increasing the thickness of the Au-Ni coating or application of an effective diffusion barrier between the coating interfaces are worthy of study in the future.

## Acknowledgment

This work was set up with a funding support of ITER Organization (SSA-50 CONV-AIF-2015-4-8). Part of the work was support by National Natural Science Foundation of China (Grant No. 11375233). The authors would like to show thanks to the electroplating engineers from Radiall® for their help towards the samples manufacturing. The views and opinions expressed herein do not necessarily reflect those of the ITER Organization.